\documentclass[12pt]{spieman}  
\usepackage{tocloft}
\usepackage{amsmath,amsfonts,amssymb}
\usepackage{mathabx}
\usepackage{graphicx}
\usepackage{setspace}
\usepackage{float}
\usepackage{hhline}
\usepackage{lineno}
\usepackage[square,numbers]{natbib}

\usepackage{caption}
\usepackage{subcaption}

\usepackage[version=4]{mhchem}
\usepackage{enumitem}

\definecolor{background}{RGB}{252, 229, 241}
\definecolor{frame}{RGB}{225, 34, 138}

\newcommand{\RNum}[1]{\uppercase\expandafter{\romannumeral #1\relax}}

\DeclareSymbolFont{UPM}{U}{eur}{m}{n}
\DeclareMathSymbol{\umu}{0}{UPM}{"16}
\let\oldumu=\umu
\renewcommand\umu{\ifmmode\oldumu\else\math{\oldumu}\fi}
\newcommand\micro{\umu}
\def\jmicron{$\micro \textrm{m}$}
\let\microns \jmicron

\title{Modeling the Impact of Starspot Inhomogeneity on Spectroscopic Retrievals of Directly-Imaged Planets}

\author[a,b,c,d]{Alex Howe}
\affil[a]{The Catholic University of America, 620 Michigan Ave., N.E. Washington, DC 20064}
\affil[b]{NASA Goddard Space Flight Center, 8800 Greenbelt Rd, Greenbelt, MD 20771, USA}
\affil[c]{Center for Research and Exploration in Space Science and Technology, NASA/GSFC, Greenbelt, MD 20771}
\affil[d]{Sellers Exoplanets Environment Collaboration, 8800 Greenbelt Road, Greenbelt, MD 20771, USA}

\author[d,e]{Natasha Latouf}
\affil[e]{NASA Postdoctoral Program Fellow, NASA Goddard Space Flight Center, 8800 Greenbelt Road, Greenbelt, MD 20771, USA}

\author[b,d]{Chris Stark}

\author[b,d]{Avi M. Mandell}

\author[b,f]{Veselin B. Kostov}
\affil[f]{SETI Institute, 189 Bernardo Ave, Suite 200, Mountain View, CA 94043, USA}

\begin{document}
\maketitle

\begin{abstract}

Stellar activity is a major complication in the detection and characterization of exoplanets by both radial velocities and transits, and the upcoming Habitable Worlds Observatory (HWO) invites us to also consider its effect on direct imaging. Spectra of directly-imaged planets can vary with the activity of their host stars because the face of the star we see is not the same as the face reflected by the planet. This discrepancy could potentially result in inaccurate measurements of the planet’s radius and unexpected, externally-caused variability in its contrast spectrum with the star. To assess the scientific capabilities and requirements of HWO, it is important that the magnitude of these effects be quantified. We present results of model retrievals of Earth-like exoplanets observed with an HWO-style survey, as they would appear when affected by starspots, using the ExoVista code for spectrum generation and the BARBIE code for spectroscopic retrieval. Both Solar-type stellar activity and highly active rapid rotators are considered and compared with an idealized quiescent host star. In the quiescent case, ${\rm SNR}\approx5$ is needed to detect atmospheric water vapor at 0.9 \jmicron\ and ${\rm SNR}\approx13$ at 0.74 \jmicron. We find that for Solar-type activity, the effect on retrievals will be negligible, but it could present problems for certain highly-active stars at limiting geometries. For an extreme case with a single large spot with 10\% coverage not visible to the observer, the SNR required for water detection increases to ${\rm SNR}\approx8$ at 0.9 \jmicron\ and ${\rm SNR}\approx18$ at 0.74 \jmicron. It also decreases the accuracy of the retrieved albedo, resulting in a value $\sim2/3$ of the true value. In light of these results, we estimate the impact that stellar variability and starspots may have on an HWO-style survey.

\end{abstract}


\section{Introduction}
\label{sec:intro}


Starspots and stellar activity present a problematic source of noise for exoplanet studies. Various forms of stellar activity can introduce noise into radial velocity (RV) time series, and stellar pulsations can introduce an RV-like oscillation directly \cite[e.g.][]{desort07,simola22}. Starspots can also cause photometric variability of a similar or greater magnitude than planets' transit signals. The discovery that our Sun is a quieter star than average (and thus, target stars are more active than expected) was a significant obstacle to the \textit{Kepler} mission \cite{basri13}. Even though starspots and transiting planets exhibit different theoretical light-curves, starspots can mimic transiting planets under some circumstances \cite{wilson12}. Further, exoplanets that transit across starspots can induce false positives in transit timing variations \cite{mazeh15}. All of these are challenges to even the first step of the detection of exoplanets.

Characterizing the properties of exoplanets in the presence of stellar activity carries additional degrees of complexity. For transiting planets, in addition to introducing photometric uncertainty into the star's baseline spectrum (and thus uncertainty in precision radius measurements), starspots have their own spectra that can mimic that of a planetary atmosphere. While many of the molecular features that occur in starspots are different from what we would expect in potentially habitable planet's atmosphere, such as TiO and VO \cite{Espinoza2019}, other molecules are potential contaminants for the planetary spectrum, most notably H$_2$O \cite{Moran2023,FT2025}. Meanwhile, as the starspots are cooler than the surrounding star, they introduce a greater deficit of blue light than red light into the spectrum, mimicking a Rayleigh slope in a planetary atmosphere in the combined transit spectrum \cite{mccullough14}. Likewise, faculae, being brighter and hotter spots on the star, would have the opposite effect, leading to an overestimate of planetary radii and a redder apparent transit spectrum.

The effect of stellar activity on exoplanet characterization depends on the specific parameters of the observation, such as the spectral range. For near-infrared (NIR) spectra such as those taken by the \textit{James Webb Space Telescope} (JWST), some models predict that the starspot effects will often be below the measurement precision \cite{zellem17}. However, in other parts of the parameter space, they may be significant to the point of needing improved stellar models to accurately fit the stellar continuum \cite{rackham24}, and actual observations of transiting planets have readily detected starspot crossing events \cite{FT2025} and apparent faculae \cite{Espinoza2025}.


The effect of stellar activity on NUV, visible, and NIR-wavelength direct imaging observations (i.e. reflection spectra), such as those planned for the Habitable Worlds Observatory (HWO), has not yet been studied in depth. Simulations of coronagraphic observations often treat stars as homogeneous point sources. There are many ways in which stellar activity \emph{could} affect coronagraphic observations. For example, flares could make certain frames unusable, quasi-resolved asymmetric stellar surfaces could modify the speckle pattern and impact the achievable noise floor, and the non-homogeneous surface could affect our ability to remove the stellar spectrum from the planet's reflected light. In this paper we focus on one piece of this puzzle: the effects of stellar activity on the retrieved planetary spectrum in the absence of chromatic stellar speckles. In other words, we investigate the astrophysical limits of stellar activity without invoking the details of the instrument's response.




Stellar activity, especially starspots, can distort the inferred reflection spectrum of a planet because the face of the star seen by the planet is not the same as the face seen by the observer. (In the ideal case of observation at quadrature, these are offset by 90 degrees). There could therefore be starspots reflected by the planet that are not visible to the observer, or vice versa, and for the most extreme activity, this could create a difference in stellar flux between the two directions of $>10$\% \cite{roettenbacher11}, enough to affect a precision retrieval. To first order, starspots decrease the continuum flux from the star, so that a starspot reflected by the planet and not visible to the observer would decrease the flux observed from the planet in an achromatic way. This would contribute an additional degeneracy to the total flux, which is already degenerate with the planet's radius, phase angle, and other potential factors such as surface roughness. However, second-order effects may also be important, especially for highly active stars. The second-order effects on reflection spectra derive from the difference in spectral shape between the star's photospheric spectrum and spot spectra, which will also affect the shape of the planet's reflection spectrum. Crucially, because starspots are cool enough for \ce{H2O} molecules to form \cite[see e.g.][]{Deming2025}, this difference in shape could change the relative depth of \ce{H2O} absorption features in the planet's spectrum, skewing the inferred abundance of \ce{H2O}. It is therefore important to understand and quantify these second order effects and the impact they might have on surveys to search for potentially habitable planets.

In this paper, we quantify the difference in retrieved \ce{H2O} abundance in the atmospheres of Earth-like planets orbiting host stars of different spectral types and levels of stellar activity, where the starspots are either less visible or unseen to the observer. We simulate planetary systems with realistic stellar models using the ExoVista code \cite{stark22,howe24}, then use the resulting planetary reflection spectra for atmospheric retrievals via the BARBIE framework and the KEN grids \cite{barbie1,barbie2,barbie3}. Our stellar models are chosen to exemplify the worst-case observational scenarios for the full range of stars that are likely to be encountered in an HWO-like target list.

In Section \ref{sec:methods}, we describe our computational methods, including updates to ExoVista to incorporate starspot modeling, an overview of the BARBIE framework, and the interface between BARBIE and ExoVista. In Section \ref{sec:models}, we describe the specific stellar and planetary models used in this study. We present the results of our nested-sampling grid-based retrievals in Section \ref{sec:results}, and we discuss the implications for various aspects of an HWO-style survey in Section \ref{sec:discuss}. Finally, we summarize our conclusions in Section \ref{sec:conclusion}.

\section{Methods}
\label{sec:methods}


\subsection{Implementing Stellar Activity in ExoVista}

The ExoVista code rapidly produces randomized simulations of planetary systems. Detailed descriptions of ExoVista can be found in \cite{stark22} and \cite{howe24}. Briefly, ExoVista assigns planets to known nearby stars by drawing from measured occurrence rates, performs stability checks, assigns each planet a classification and wavelength-dependent albedo based on their location in parameter space, integrates the systems' equations of motion using an $n$-body integrator, and generates a scattered light model of a debris disk consistent with the underlying planetary system. ExoVista also computes stellar radial velocities, astrometric positions, and transit depths, as well as all of the inputs required for direct imaging simulations.

In this paper we describe an updated version of the software, ExoVista 2.5. ExoVista 2.5 incorporates stellar activity for the first time, including both starspots and granulation. Starspots in the stellar model are generated according to the method of VSPEC \cite{vspec}, adapted for integration into ExoVista. In this method, each starspot and its properties may be directly specified or randomly generated, and each spot has its own location, size, and area ratio of penumbra to umbra. (The spots are assumed to be circular.) ExoVista supports time evolution of starspots, although we use only the initial spot configurations for each case in this paper. In the time-evolution mode, each spot has its own growth rate, maximum size, and decay rate specified, and new spots are added as needed to keep the total spot area approximately constant.



Globally, each star has four starspot parameters: umbra temperature, penumbra temperature, spot coverage as a fraction of surface area, and the spot distribution pattern. Spots are allowed to overlap, so that in the limit of very high spot coverage, the spot coverage fraction behaves as an optical depth. (That is, a star with a nominal 100\% spot coverage would have a true coverage on average of $1-\exp(1.00)=63\%$.) However, this will be only a minor factor for a realistic spot coverage of $\le0.1$. Spot areas are measured in units of micro-solar hemispheres (MSH). Three distribution patterns may be selected: a random distribution over the entire star, a solar-like distribution (with the spots being concentrated in two belts in the temperate latitudes), and a single large spot covering the requisite area.

Additionally, if the initial configuration of spots is not in an equilibrium state with the correct starspot coverage for any reason, it is also possible to time-evolve the star forward until it reaches a suitable equilibrium before beginning the orbital integration and spectrum calculations. The timescale for this is identified as the ``warm-up time'' and may also be varied.

Granulation is computed by assuming a fixed percentage of the star's non-spot area is at a slightly lower effective temperature than the rest of the photosphere by $\Delta T$ in the lanes between the granules. Both $\Delta t$ and the coverage of the lowered temperature may be varied. However, they are not treated as time-varying over the code's integration. Due to the high frequency of fluctuations in granules, this component of variability is assumed to average out over typical observation timescales.

Our adopted values of $f_{\rm gran}=0.2$ and $\Delta T=-200 K$ for granulation are most consistent with M-dwarfs, while earlier spectral types may have a $\Delta T$ that is larger in magnitude \cite{magic2014}. However, we can expect that our results will not be sensitive to the choice of granulation parameters because we obtain similar results for G-dwarf, K-dwarf, and M-dwarf host stars, which have far more divergent effective temperatures.

The default values for the starspot parameters are listed in Table \ref{tab:starspots}.

\begin{table}[htb]
    \centering
    \begin{tabular}{l | r }
    \hline
    Spot Parameters \\
    \hline
    Warmup time (d) & 0.0 \\
    Initial area (MSH) & 10 \\
    Growth rate (d$^{-1}$) & 0.52 \\
    Decay rate (MSH d$^{-1}$) & 10.89 \\
    \hline
    Spot Hyperparameters \\
    \hline
    Mean maximum area (MSH) & 500 \\
    Mean penumbra-to-umbra ratio & 5 \\
    Log-Sigma area & 0.2 \\
    \hline
    Global Parameters \\
    \hline
    Spot distribution & random \\
    Spot coverage & 0.2 \\
    Penumbra $T_{\rm eff}$ (K) & 2700 \\
    Umbra $T_{\rm eff}$ (K) & 2500 \\
    Granule lanes coverage & 0.2 \\
    Granule lanes $\Delta t$ (K) & -200 \\
    \hline
    \end{tabular}
    \caption{Star and starspot models. The growth rate is exponential, and the decay rate is linear. Granulation coverage applies to the spotless fraction of the star's surface.}
    \label{tab:starspots}
\end{table}

For the purpose of computing stellar spectra, the geometry of each starspot is simplified and represented as a circular umbra centered in a circular penumbra, both modeled as flat disks and arranged in three dimensional space such that the circumferences of both the umbra and penumbra lie on the surface of the star. (Thus, the umbra is placed slightly above the penumbra.) This arrangement is not quite accurate for realistic starspots, but we use it to simplify and speed up the calculations of the projected spot areas compared with a pixel-based approach. We also use the same umbra and penumbra temperatures of 2500 K and 2700 K, respectively across spots and spectral types, even for late-type stars. This is in part to avoid edge effects from the lower temperature limit of 2300 K in the PHOENIX model grid. Because the flux from spots is much lower than that from the quiescent star, temperature variation between spots will have a negligible effect.

In ExoVista 2.5, the stellar spectrum models of \cite{Kurucz03} have been replaced with a set of spectra from the PHOENIX models, specifically the ``R10000FITS PHOENIX-ACES-AGSS-COND-2011'' HiResFITS model set with Solar abundances \cite{Husser13}. This grid provides stellar spectra over a range in $T_{\rm eff}$ from 2300 K to 12000 K and $\log g$ from 0.0 to 6.0. As provided, the spectral subgrid covers the range of 300-10000 nm at a resolution of 0.1 nm. Thus, the grid provides updated model spectra over a wider wavelength range, and also extends to lower effective temperatures than the Kurucz models.

The total stellar spectrum is computed as a linear combination of spectra from the PHOENIX models \cite{Husser13} at the quiescent photosphere, quiescent granule lanes, penumbra, and umbra temperatures of the star, with the weights depending on the projected area of each component as seen by the observer or by the planet, as appropriate. The projected spot area is computed analytically by projecting the spots onto the stellar disk, tilting each spot according to its phase angle relative to the line of sight. This also involves some simplifying geometric assumptions. A foreshortened circle on a spherical surface will not be a perfect ellipse; a spot viewed almost exactly edge on may be not counted or may be double-counted because the umbra is at a different altitude than the penumbra; and for the stellar spectra as seen by the planet, the fact that the limb of the star is not perfectly perpendicular to the line of sight is neglected. However, all of these artifacts apply only to spots that are viewed nearly edge on, so that they will contribute much less to the total projected spot area. Thus, the error in computed spot coverage is $\ll1\%$ of the desired coverage.

Finally, limb darkening is applied to the stellar disk using a quadratic limb darkening model \cite{Kopal1950} of the form:
\begin{equation}
    f=1-(u_1+1)\mu - u_2\mu^2,
\end{equation}
where we adopt $u_1=0.0473$ and $u_2=0.0841$, based on the values for the Sun in \cite{Claret2013}.

A single limb darkening fraction is computed for each spot based on its phase angle relative to the line of sight, while the quiescent part of the star has a limb darkening factor averaged over the entire disk of the star. Again, this neglects second-order geometric effects such as the non-zero extent of the spots and the uneven area distribution of the quiescent photosphere, and spectral variation in the limb darkening curve. The error due to the non-zero extent of the spots can be bounded by the error between the quadratic limb darkening law and a linear one. This proves to be $\frac{e_f}{f}<1-\frac{f(1)-f(0)}{f(0.5)}=0.046$ for our adopted coefficients. The spectral variation can be somewhat larger. For example, substituting limb darkening coefficients for TRAPPIST-1 \cite{Claret2012} as a proxy for the lower spot temperature yields $\frac{e_f}{f}\sim0.25$ for an impact parameter of $b=0.7$ compared with the Solar values. However, these errors will necessarily remain smaller than the effect of spot area, and their contribution to the overall stellar flux will be greatly reduced due to the lower flux from the spots.

In this study we make use of three stellar activity models for each of three stellar spectral types we select based on this library, as described in Section \ref{sec:models}. Future developments of ExoVista will expand on this stellar activity model by implementing stellar rotation, time evolution of starspots, and the addition of faculae.

ExoVista 2.5 contains a number of other changes beyond stellar activity, which are not used in this work, including low-resolution (0.01 days) transit and eclipse detection, and a faster $n$-body integrator translated from Python to C++ to facilitate transit detection. It also includes functionality to convert output files back to input files for iteration purposes and functionality to directly control more exozodiacal disk variables in the system input files. And in the planetary system model itself, two new components of random variation have been implemented, each of which may be optionally added to the system generation to better replicate the observed distribution of exoplanets. A random variation may be added to the planetary radii based on the uncertainty ranges in the mass-radius relation of \cite{ChenKipping}. Meanwhile, a random variation may be added to the Lambertian phase function, which allows it to vary smoothly within the range of phase functions observed for Solar System bodies.

Additionally, in these updates, relevant data were added to the output FITS files for more flexible post-processing, including the phase angles of planets. For the use case of generating a universe of random planetary systems, the function to assign planets to stars was also rewritten to be more faithful to the observed occurrence rates while including natural variation in the population. For this project, we ran ExoVista on a single Intel Core i5-8279U processor, and with typical inputs, scene generation took 5-10 minutes per star.

\subsection{Creation of Exoplanet Spectra}

For our fiducial exoplanet, we adopt the wavelength-dependent albedo of a modern Earth twin with an isotropic atmosphere, created using the Planetary Spectrum Generator \cite[PSG,][]{PSG,PSGbook}. The assumed volume mixing ratios (VMRs) for \ce{H2O}, \ce{O2}, and \ce{O3} are $3\times10^{-3}$, $0.21$, and $7\times10^{-7}$, respectively. We adopt a constant temperature profile of 250 K, an albedo ($\mathrm{A_{s}}$) of 0.3, pressure ($\mathrm{P_{0}}$) of 1 bar, and a planetary radius of $\mathrm{R_p}$ = 1 $\mathrm{R_\Earth}$ \cite{feng18}. The model also assumes a cloud coverage of 0\%, in order to completely isolate the effects of the starspots from planetary variation. 

We import the albedo derived from PSG into ExoVista. ExoVista then places the planet at quadrature and illuminates the planet with light from the spotted star given the stellar parameters and orientations described in Section \ref{sec:models}. Assuming a Lambertian phase function, ExoVista self-consistently calculates the planet's observed contrast spectrum between its reflection spectrum (reflecting the partially-unseen planet-facing stellar spectrum) and the observer-facing stellar spectrum. The stellar spectrum is approximated based on an orthographic projection of the stellar disk, for which any errors will be small at the distance of the habitable zone. The spectra are computed at the native resolution of the PHOENIX models, then binned to the desired resolution (in this case $R=140$, which is the standard baseline assumption for HWO \cite{luvoir}), by a boxcar binning method. 

\subsection{Bayesian Spectral Retrievals}

In order to understand how the starspots affect the strength of molecular detections, we run a series of spectral retrievals on the simulated contrast spectra generated by ExoVista. For this we use the Bayesian Analysis for Remote Biosignature Identification on exoEarths (BARBIE) methodology, similar to prior studies by \cite[hereafter BARBIE1, BARBIE2, and BARBIE3, respectively]{barbie1, barbie2, barbie3}. We make use of the KEN grid created and validated in BARBIE3, which contains millions of pre-computed spectra covering a large, predefined parameter space for a range of molecules. Here we specifically utilize the Merman KEN grid set, which contains \ce{H2O}, \ce{O2}, and \ce{O3}. We bin the spectra within the KEN grid from the native resolving power $R=500$ to $R=140$, also using a boxcar binning method. We examined a range of assumed signal-to-noise ratios (SNRs) for the data set. The noise term for the SNR herein refers to a standard deviation of noise at each pixel, and is added as Gaussian noise after the spectrum is simulated to mimic the SNR in real observations to first-order. We are not using a random Poisson draw, as the results will not vary significantly due to noise without a pathological noise case, which is presently not constrained for HWO.

We run a set of Bayesian spectral retrievals using the nested sampling tool PSGnest, which is housed within PSG.  PSGnest is adapted from the original Fortran version of the Multinest retrieval algorithm \cite{multinest} and has been created specifically to work with grids to decrease the computation effort required for retrievals. We use the log-evidence (LogZ) value calculated by the retrieval process \cite{PSGbook} to calculate the log-Bayes factor \cite[$\mathrm{lnB}$;][]{benneke13}, which directly compares the retrievals with and without \ce{H2O} and estimates the likelihood of the existence of \ce{H2O} in the exoplanet's atmosphere. We also compute a 68\% confidence interval for each retrieved parameter \cite{harrington22} in our retrievals in order to determine the accuracy of each detection. These results are shown in the 1D histograms of Figures \ref{fig:corner} and \ref{fig:corneractive}. Following BARBIE1, we vary our metrics slightly from what is described in Table 2 of \cite{benneke13}; in our metrics,  $\mathrm{lnB}<2.5$ is unconstrained (i.e. no detection, as opposed to a weak detection), $2.5\le\mathrm{lnB}<5.0$ is a weak detection (as opposed to a moderate detection), and $\mathrm{lnB}{\ge}5.0$ is a strong detection. We vary these definitions as, when following \cite{benneke13}, a weak detection would have an extremely broad 68\% credible region, and result in no constraint on the parameter. By altering the definitions slightly, a weak detection would be slightly constrained. These $\mathrm{lnB}$ values for the results of our retrievals are shown as the heat maps in Figure \ref{fig:heatmaps_h2o}.

\section{Model Selection for Stars and Starspots}
\label{sec:models}

We select a total of nine stellar models in this study, spanning the likely parameter space of stellar activity for the target stars in HWO's exoEarth survey. Specifically, we select three host star spectral types, G2V, K5V, and M2V, each with three different starspot models applied. 
Our stellar models use Solar mass, radius, and luminosity for the G2 case and the stellar values adopted by \cite{MamajekII} for the K5 and M2 cases.

For each spectral type, we apply three stellar activity models: Quiet, Active, and Extreme. The Quiet model has granulation and limb darkening, but no starspots. The Active model adds a solar-type starspot distribution with a spot coverage of 1\%. This is somewhat higher than the largest spot coverage recorded for the Sun at 0.5\% \cite{hathaway15}, although it may be typical of field G-dwarfs, which tend to be more active than the Sun on average \cite{basri13}. As discussed in Section \ref{sec:design}, it appears to be near the 95th percentile of activity of expected target stars in the \textit{Kepler} survey, so it will serve as a useful reference point. The Extreme model features a single large spot with a coverage of 10\%, placed just behind the limb of the star as seen by the observer. A single spot or a small number of very large spots covering $\sim$10\% of the star's surface has been inferred from light curves of some young rapid rotators \cite[see e.g.][]{roettenbacher11}; this type of activity may be especially important for low-mass host stars, which are more active than earlier spectral types.

The parameters of the host star and starspot models used in this paper are provided in Table \ref{tab:starmodels}.

\begin{table}[htb]
    \centering
    \begin{tabular}{l | r | r | r}
    \hline
    Spectral Type     & G2V & K5V & M2V \\
    \hline
    $M_V$             & 4.83 & 7.28 & 10.124 \\
    $B-V$             & 0.65 & 1.150 & 1.46 \\
    $L_*$ ($L_\odot$) & 1.0 & 0.174 & 0.0263 \\
    logg (cm/s)       & 4.40 & 4.585 & 4.80 \\
    $T_{\rm eff}$ (K) & 5778 & 4440 & 3550 \\
    Mass ($M_\odot$)  & 1.0 & 0.70 & 0.44 \\
    $R_*$ ($R_\odot$) & 1.0 & 0.701 & 0.434 \\
    \hline
    \hline
    Starspot Distribution &  Quiet    & Active  & Extreme  \\
    \hline
    Coverage & 0\% & 1\% & 10\% \\
    Distribution & None & Solar & Single Spot \\
    \hline
    \end{tabular}
    \caption{Star and starspot models.}
    \label{tab:starmodels}
\end{table}

In each case, the selected system parameters constitute a ``worst-case scenario'' from the perspective of characterization. In the Active case, we use a face-on system orientation and pole-on stellar orientation (assuming the star's rotation axis and the planet's orbit to be aligned). In this configuration, Solar-type starspots appear near the limb of the star as seen by the observer and will have a relatively small effect on the star's observed spectrum, while they will fall in a nearly face-on belt as seen by the planet. For comparison, an edge-on alignment or intermediate geometry may show low variability while remaining quite active, but this would greatly diminish the star's variability as a function of viewing angle, and thus, the planet's albedo could be measured more accurately.

\begin{figure*}[]
\centering
\includegraphics[width=0.90\textwidth]{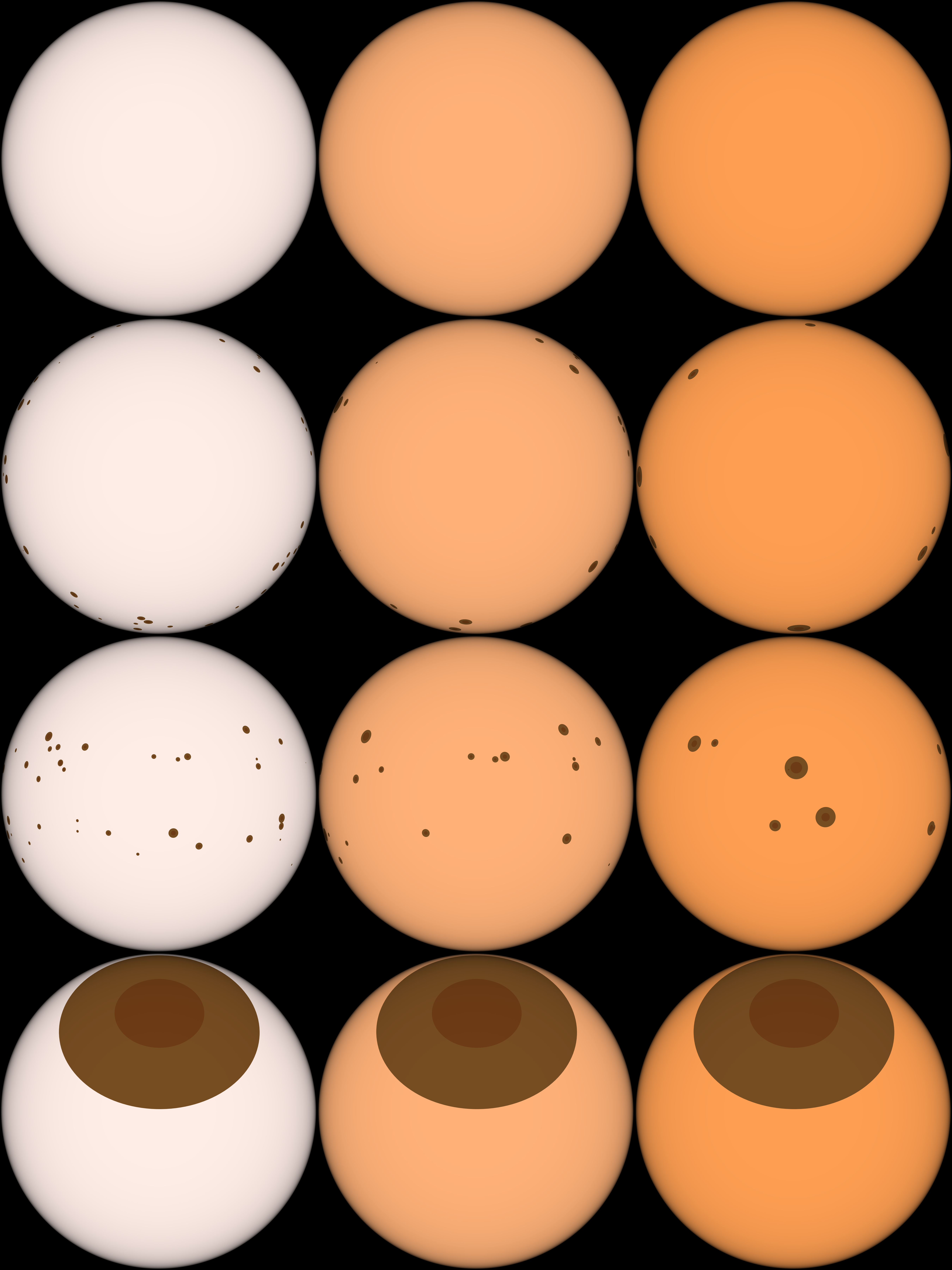}
\caption{Images of the stellar disks modeled in this study. Left column: G2 star. Middle column: K5 star. Right column: M2 star. Top row: spotless disk. Second row: Solar-type activity as seen by the observer. Third row: Solar-type activity as seen by the planet. Bottom row: extreme stellar activity as seen by the planet. The colors reflect the actual stellar colors of the selected spectral types, based on \cite{harre21}.}
\label{fig:starspots}
\end{figure*}

For the Extreme case, we again adopt a face-on system and pole-on star, with the single spot located below the equator, placing it entirely out of view of the observer, but again nearly face-on as seen by the planet. Both of these cases thus maximize the differences between the observed stellar spectrum and the reflected spectrum from the planet. The face-on orientation also ensures that all observations of the planet are at quadrature. Notably, the system orientation for the Extreme scenario not only maximizes the stellar variability, but also minimizes the detectability of that variability, whereas in most scenarios with such high starspot coverage, the stellar variability will be detectable in the star's light curve as seen by the observer. Images of the stellar disk for each starspot model as seen by the observer and by the planet are shown in Figure \ref{fig:starspots}.

Note that we have made several simplifying assumptions to our model, particularly, neglecting time-variability. The same star may have different activity levels at different times, and a star that is ``quiescent'' in terms of spot coverage may become ``active'' in a later observation, or \textit{vice versa}. For the purposes of this study, such variability is not relevant because we are interested in the worst-case scenarios of a maximum deviation from a featureless star, which requires only that a star be quiescent or active at a particular snapshot in time. Our results show that for most stellar targets, any such variability is too small to have a statistically significant effect regardless, but future direct imaging studies involving highly active stars should account for potential changes in the stellar spectrum.




\section{Results}
\label{sec:results}


\begin{figure*}[]
\centering
\includegraphics[width=0.99\textwidth]{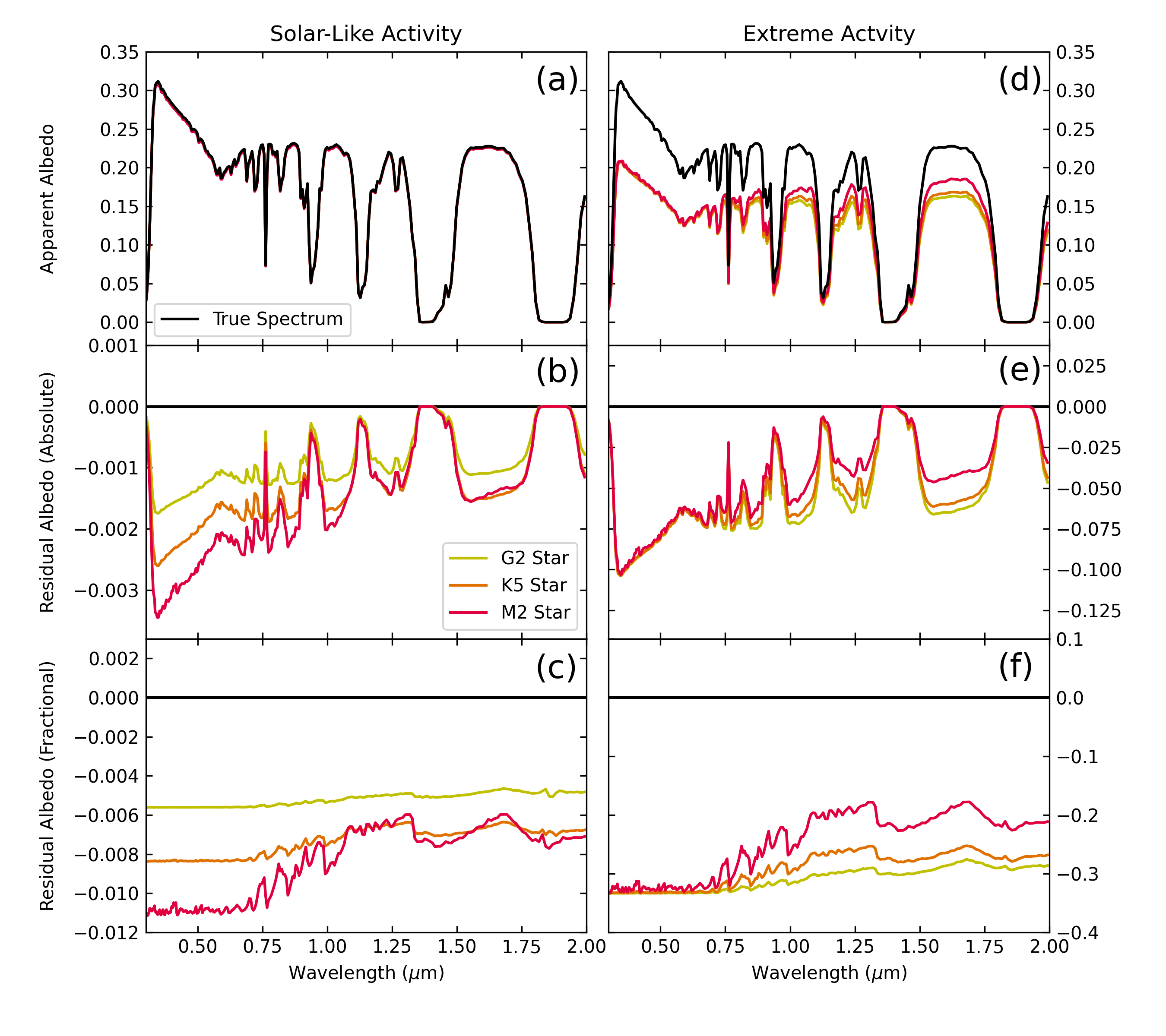}
\caption{Inferred planetary albedo spectra and residuals for the observations modeled in this study. Panels (a,b,c): solar-type activity. Panels (d,e,f): extreme activity. Panels (a,d): the true albedo spectrum (black) and the inferred albedo spectra for the simulated observations. (The true spectrum also corresponds to the observations in the ``quiet'' scenario.) Panels (b,e): absolute residuals of the inferred spectrum minus the true spectrum. Panels (c,f): fractional residuals of the inferred spectrum divided by the true spectrum, minus one. For solar-type activity, the major source of error in the expected continuum is the number of spots seen by the planet. For extreme activity, the major source of error is limb darkening across the width of the spot, which is calculated only for the spot center.}
\label{fig:diffspectra}
\end{figure*}


    
Figure~\ref{fig:diffspectra} portrays how varying the stellar type and activity level influences the inferred planetary geometric albedo spectrum. The black lines correspond to the true spectrum as well as the inferred spectrum for a quiet star, while the colored lines are the inferred albedo spectra with different stellar activity models. The difference between a quiet star (i.e. no starspots) and an active star is indistinguishable by eye. Thus, we can conclude that any difference between a quiet star and Solar-like activity will not be seen in a retrieved albedo spectrum. The extremely active stars, however, produce a fainter inferred spectrum, mimicking a lower geometric albedo. Notably, the location of the depth of the feature does not change (so the deepest point in the feature remains at the same albedo value regardless of the change in continuum), i.e. in an extremely active spectrum, the molecular features are artificially shallower due to the effect of starspots.

\begin{figure}
    \begin{subfigure}[t]{0.49\textwidth}
        \centering
        \includegraphics[width=\linewidth]{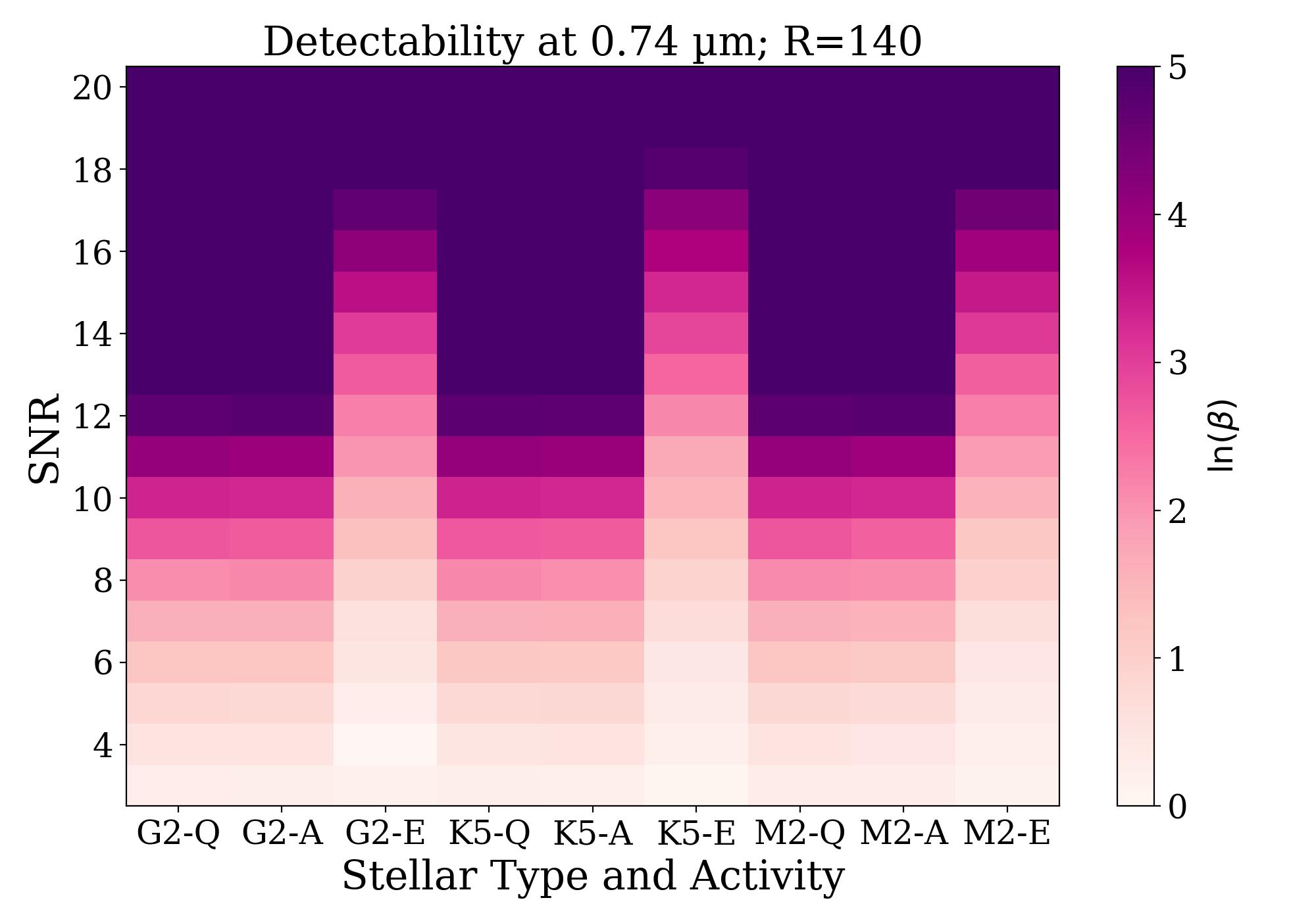}
        \caption{Detectability of \ce{H2O} for the 0.74 {\microns} water feature.}
    \end{subfigure}
    \hfill
    \begin{subfigure}[t]{0.49\textwidth}
        \centering
        \includegraphics[width=\linewidth]{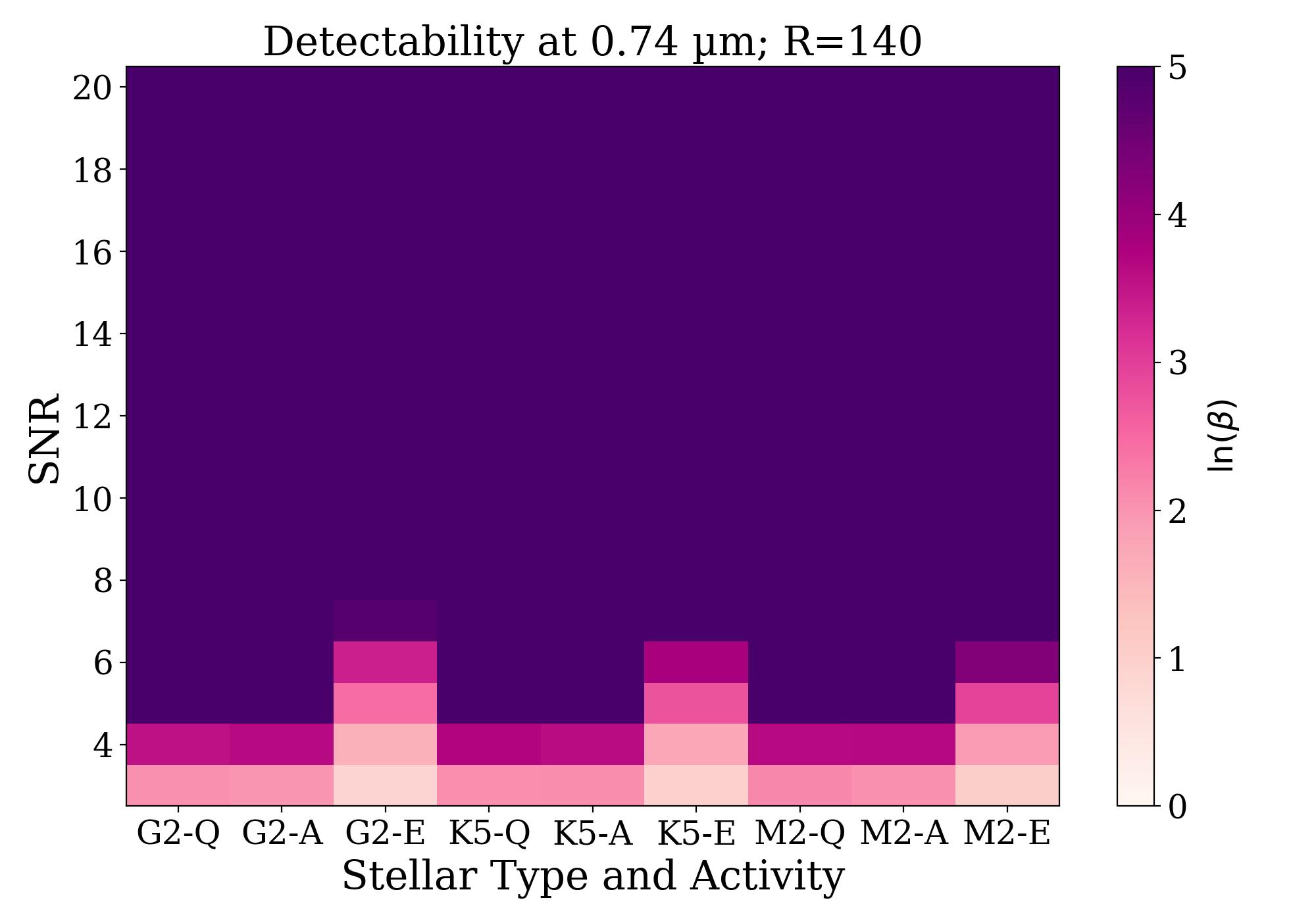}
        \caption{Detectability of \ce{H2O} for the 0.9 {\microns}}
    \end{subfigure}
\caption{Heatmap plots illustrating detection strength as a function of SNR and varying stellar type and activity. SNR is on the y-axis, stellar type and activity level is on the x-axis (where Q is quiet, A is active, and E is extreme), and the color bar shows the range of log-Bayes Factor ($\ln B$) from 0 to 5 to describe detection strength. $\ln B<2.5$ are unconstrained, $2.5 \leq \ln B < 5$ are weak, and $\ln B>5$ are strong.}
\label{fig:heatmaps_h2o}
\end{figure}

Figure~\ref{fig:heatmaps_h2o} presents two heatmaps summarizing the \ce{H2O} detectability assuming a modern Earth abundance (3$\times10^{-3}$ VMR) as a function of SNR and stellar type and activity. Both color bars cover the full range of results between 0 and 5 log-Bayes factor, to represent the change in detection strength from unconstrained to strong. Figure~\ref{fig:heatmaps_h2o}a shows the detectability at the 0.74 {\microns} \ce{H2O} feature, and Figure~\ref{fig:heatmaps_h2o}b shows the detectability at the 0.9 {\microns} \ce{H2O} feature. Figure~\ref{fig:heatmaps_h2o} illustrates that at 0.9 {\microns} and 0.74 {\microns}, \ce{H2O} is detectable regardless of the stellar activity level depending on the SNR. Extreme stellar activity heavily influences the detectability of \ce{H2O} at every wavelength, resulting in increased SNRs required for strong detection compared with either the quiet or active stellar activity ($\sim$18 compared to 13). This effect is much more pronounced at shorter wavelengths where the \ce{H2O} features are shallower than at 0.9 {\microns}. 

\begin{figure*}
    \begin{subfigure}[t]{0.49\textwidth}
        \centering
        \includegraphics[width=\linewidth]{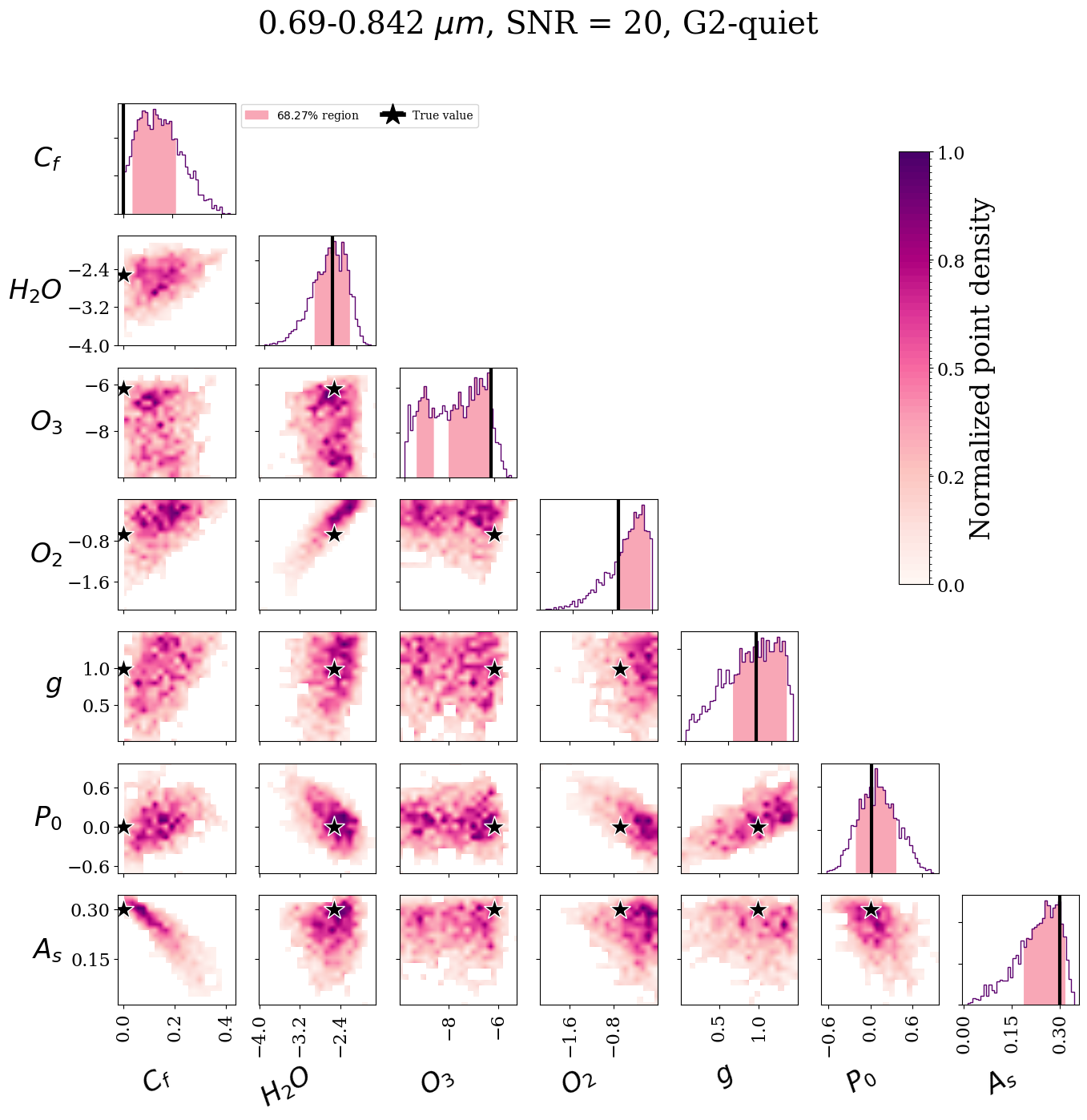}
        \caption{Corner plot for the bandpass centered on 0.74 {\microns} for a quiet G2 star.}
    \end{subfigure}
    \hfill
    \begin{subfigure}[t]{0.49\textwidth}
        \centering
        \includegraphics[width=\linewidth]{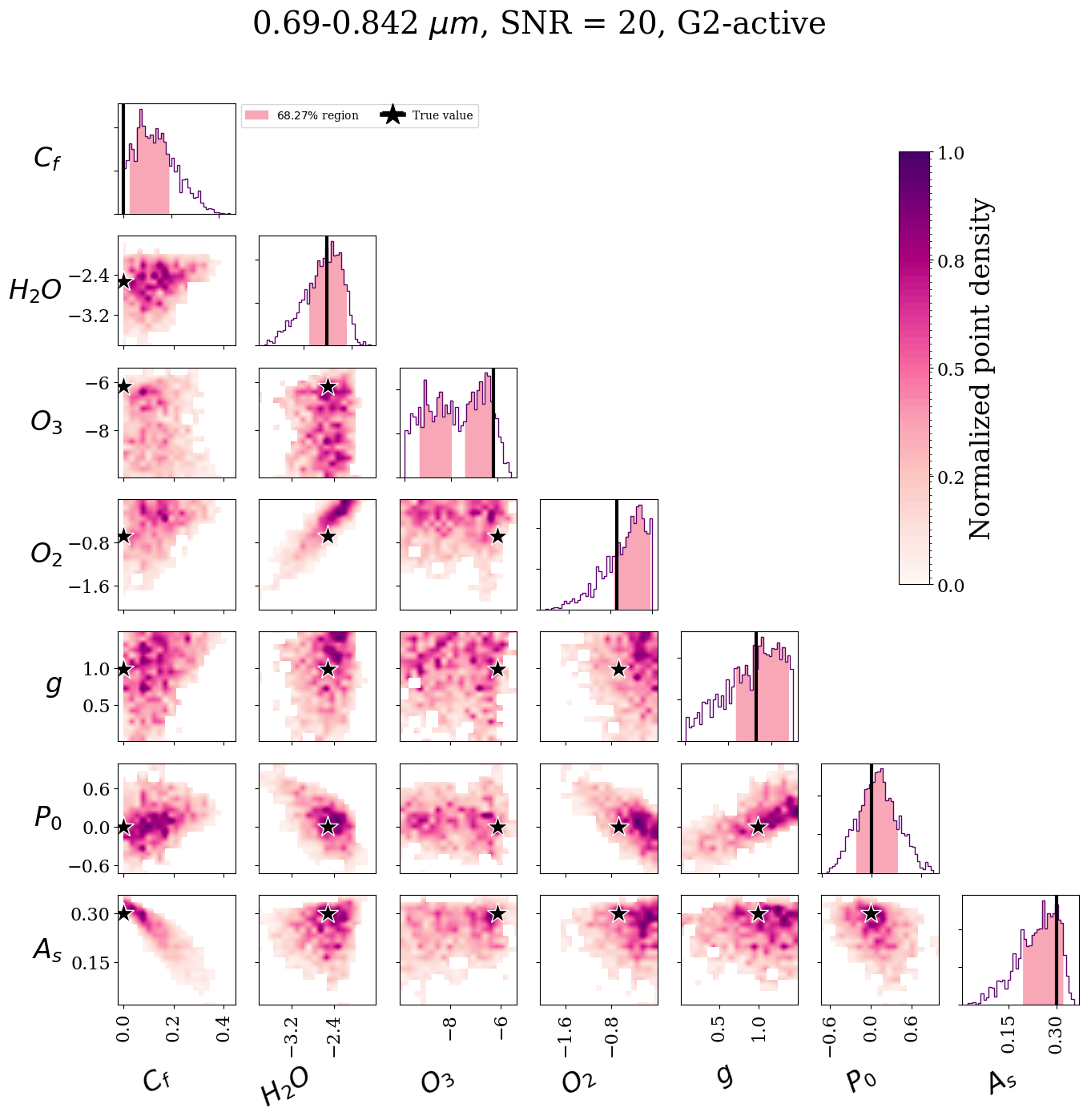}
        \caption{Corner plot for the bandpass centered on 0.74 {\microns} for an active G2 star.}
    \end{subfigure}
\caption{Two corner plots, one for a quiet G2 star and one for an active G2 star. Each has the same parameters with SNR of 20, in order to confirm that all \ce{H2O} features are detectable. In all corner plots, the 68\% credible regions are shown as pink shading in the 1D marginalized posterior distributions along the diagonal of the corner plot, and the true values are represented by black lines in the diagonals of the corner plot, and black stars within the 2D plots.}
\label{fig:corneractive}
\end{figure*}

The active and quiet stellar activity levels appear to yield effectively identical results regardless of wavelength, with any differences not significantly affecting SNR requirements. We further confirm this in Figure~\ref{fig:corneractive}. Figure~\ref{fig:corneractive} presents two corner plots investigating the 0.74 {\microns} \ce{H2O} feature for a quiet star (Figure~\ref{fig:corneractive}a) and a normally active star (Figure~\ref{fig:corneractive}b). No significant differences in the 1D marginalized posterior distribution or the 2D plots are visible. We therefore only consider the comparison of a quiet star to an extremely active star in the remainder of our results. We focus our results on the detection and constraint of \ce{H2O}. Although there are other molecules present in the VIS wavelength regime (Chappuis band \ce{O3} and \ce{O2} a-band), we leave investigation of them to a future work. However, they are included as free parameters in our retrievals and thus will appear in our corner plot results for thoroughness.

\begin{figure*}
    \begin{subfigure}[t]{0.49\textwidth}
        \centering
        \includegraphics[width=\linewidth]{snr20_all_0.69_0.842_G2-quiet.png}
        \caption{Corner plot for the bandpass centered on 0.74 {\microns} for a quiet G2 star.}
    \end{subfigure}
    \hfill
    \begin{subfigure}[t]{0.49\textwidth}
        \centering
        \includegraphics[width=\linewidth]{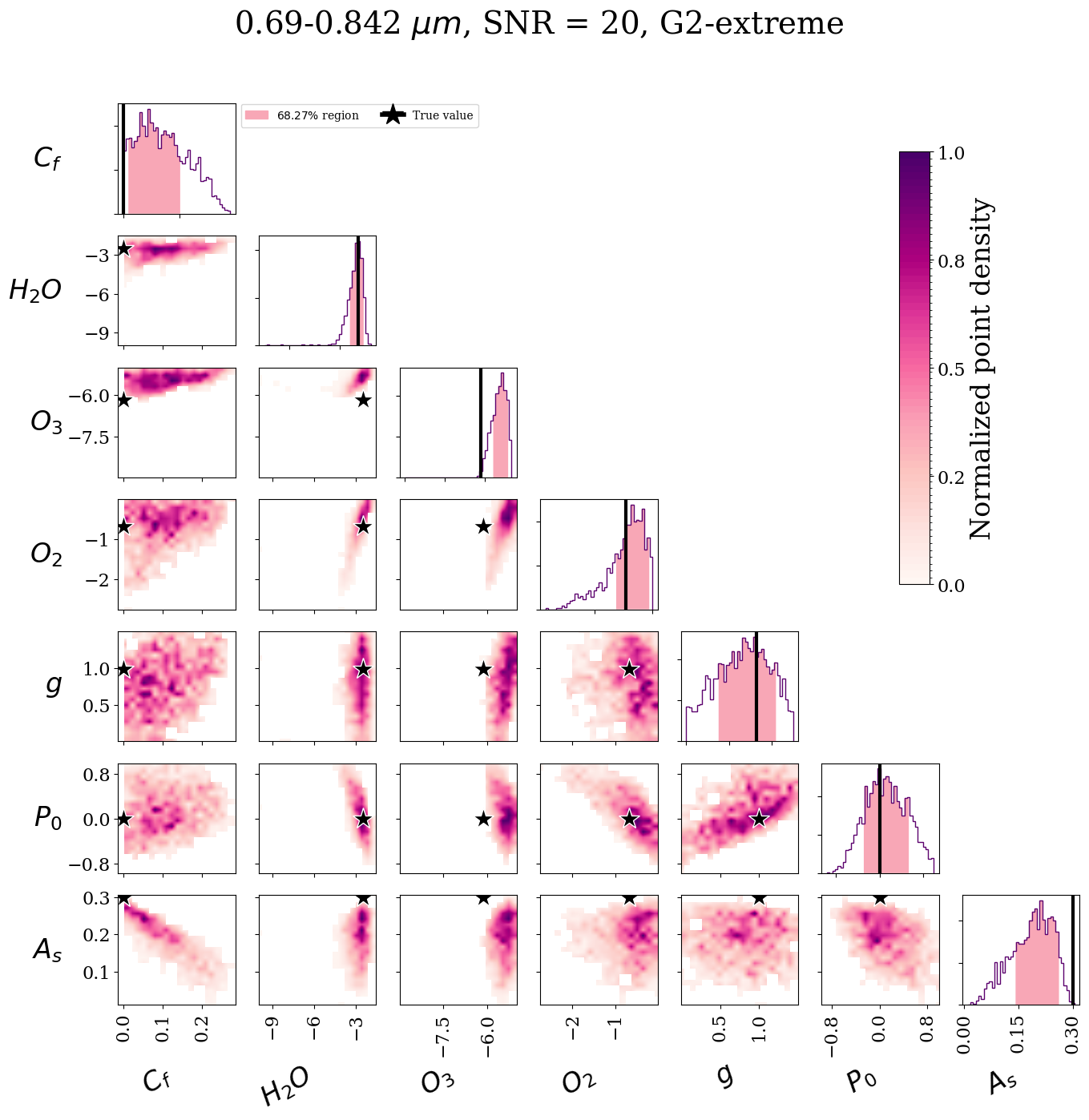}
        \caption{Corner plot for the bandpass centered on 0.74 {\microns} for an extremely active G2 star.}
    \end{subfigure}
    \vspace{1cm}
    \begin{subfigure}[t]{0.49\textwidth}
        \centering
        \includegraphics[width=\linewidth]{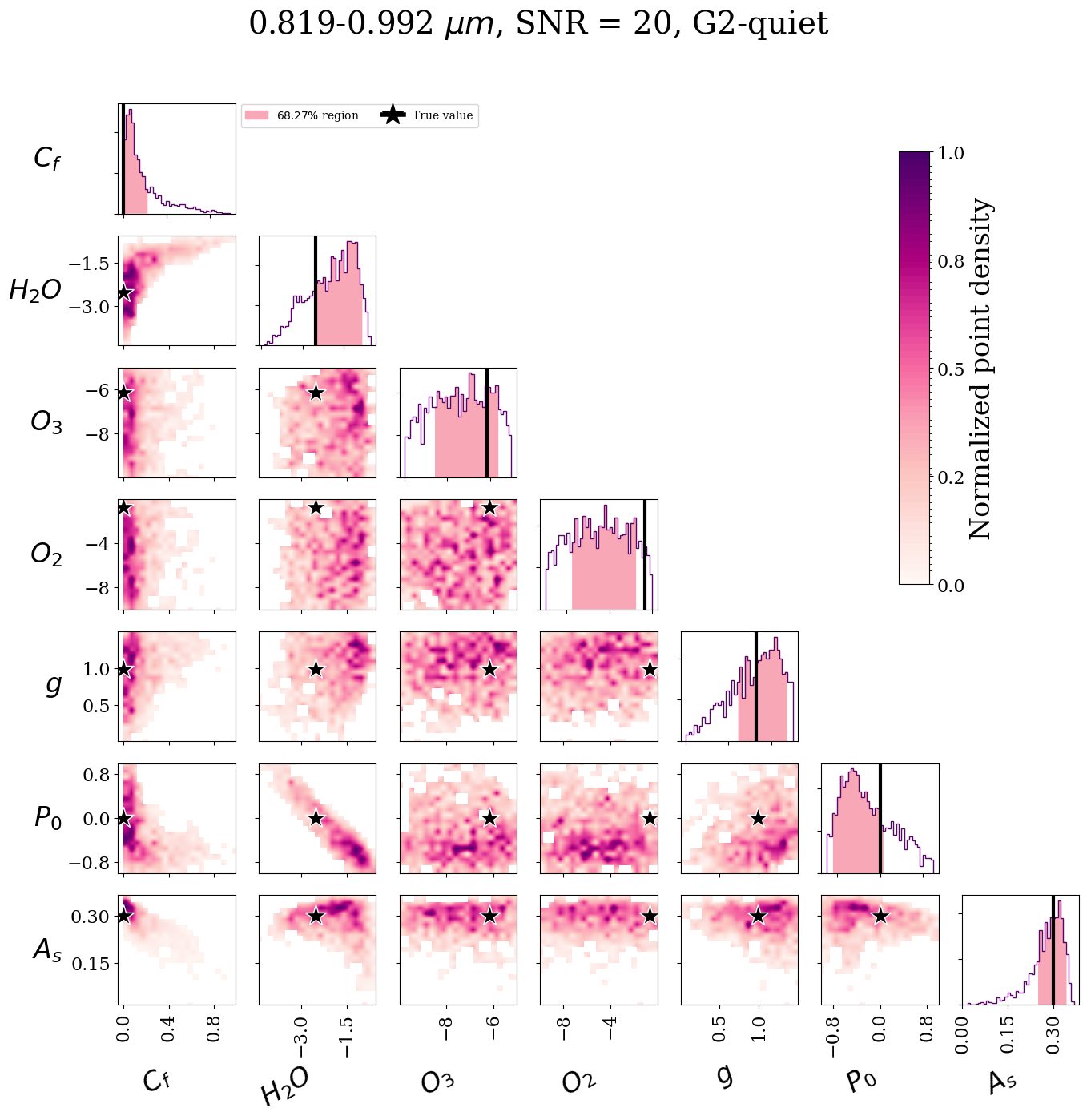}
        \caption{Corner plot for the bandpass centered on 0.9 {\microns} for a quiet G2 star.}
    \end{subfigure}
    \hfill
    \begin{subfigure}[t]{0.49\textwidth}
        \centering
        \includegraphics[width=\linewidth]{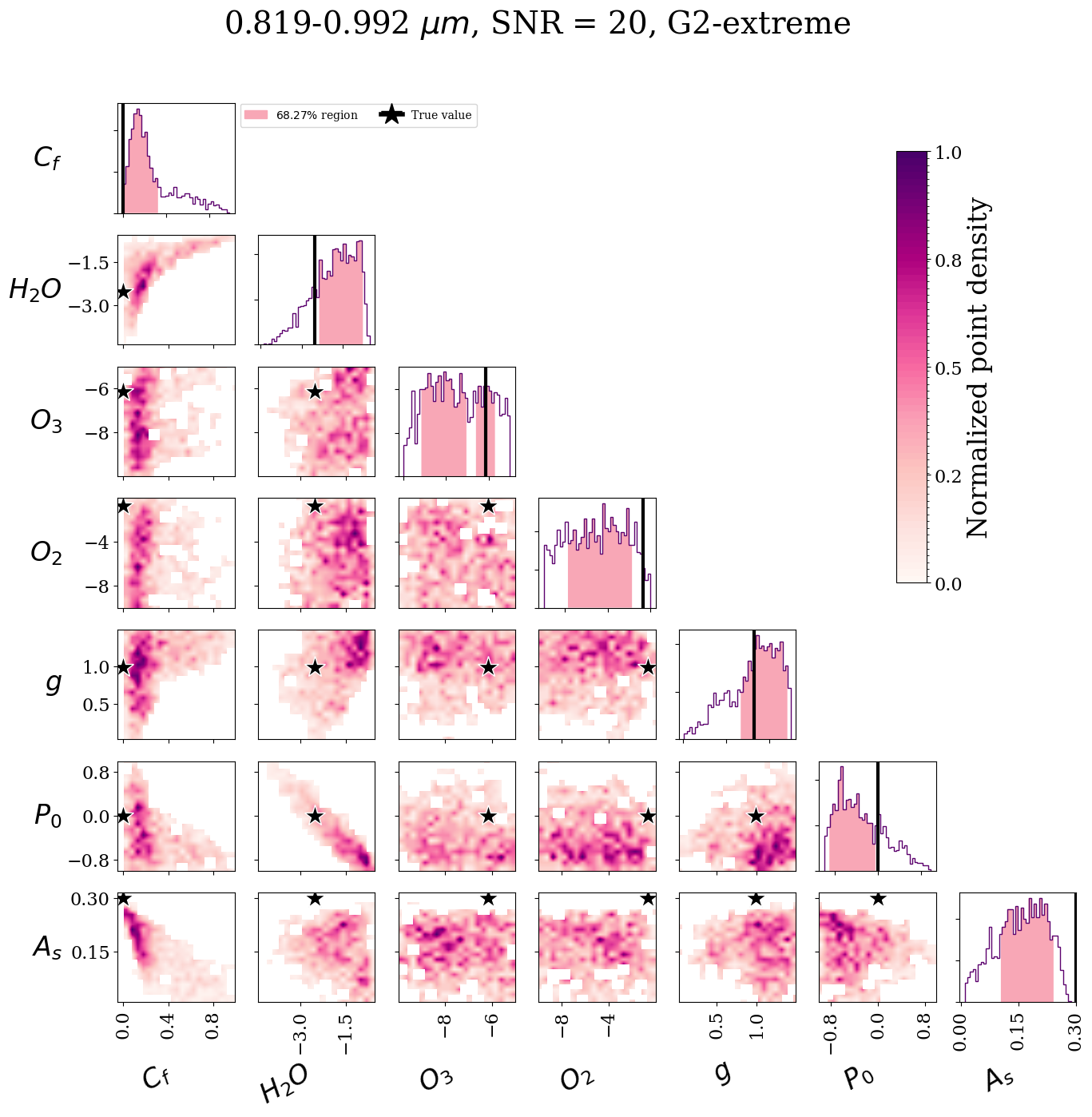}
        \caption{Corner plot for the bandpass centered on 0.9 {\microns} for an extremely active G2 star.}
    \end{subfigure}
\caption{Four corner plots, two for a quiet G2 star and two for an extremely active G2 star. All other plot aspects are identical to Figure \ref{fig:corneractive}.}
\label{fig:corner}
\end{figure*}

We expect $\mathrm{A_{s}}$ to be degenerate with stellar activity, especially with extreme stellar activity, and thus, we present another series of corner plots in Figure~\ref{fig:corner} to investigate degeneracies due to stellar activity levels.

\begin{table}[ht]
    \centering
    \hspace{-1cm}
    \begin{tabular}{lcccccc}
    \hline
    \textbf{} & $log_{10}\ce{H2O}$ & $log_{10}\ce{O3}$ & $log_{10}\ce{O2}$ & $log_{10}g$ & $log_{10}P_{0}$ & $A_{s}$ \\
    \hline
    \textbf{Truths} & \textbf{-2.52} & \textbf{-6.15} & \textbf{-0.68} & \textbf{0.99} & \textbf{0} & \textbf{0.3}\\
    \hline
    & & & 0.9 {\microns} & & &\\
    \hline
    G2-quiet & $-1.86^{+1.03}_{-0.71}$ & $-7.33^{+1.73}_{-1.25}$ & $-4.84^{+3.35}_{-2.65}$ & $1.01^{+0.41}_{-0.27}$ & $-0.32^{+0.38}_{-0.56}$ & $0.29^{+0.05}_{-0.04}$ \\
    G2-extreme & $-1.75^{+0.97}_{-0.61}$ & $-7.52^{+1.81}_{-1.64}$ & $-4.98^{+3.28}_{-2.75}$ & $1.04^{+0.41}_{-0.24}$ & $-0.39^{+0.34}_{-0.52}$ & $0.16^{+0.08}_{-0.06}$ \\
    \hline
    & & & 0.74 {\microns} & & &\\
    \hline
    G2-quiet & $-2.57^{+0.43}_{-0.34}$ & $-7.71^{+1.57}_{-1.75}$ & $-0.44^{+0.41}_{-0.25}$ & $0.95^{+0.47}_{-0.28}$ & $0.09^{+0.30}_{-0.34}$ & $0.24^{+0.08}_{-0.05}$ \\
    G2-extreme & $-2.73^{+0.61}_{-0.47}$ & $-5.44^{+0.32}_{-0.24}$ & $-0.60^{+0.52}_{-0.33}$ & $0.82^{+0.45}_{-0.36}$ & $0.11^{+0.43}_{-0.40}$ & $0.19^{+0.07}_{-0.05}$ \\
    \hline
    \end{tabular}
    \caption{Retrieved parameter values at SNR=20 for the 0.9 {\microns} and 0.74 {\microns} bandpass centers. Each cell gives the median retrieved value with the credible region envelope. The true values are provided in the top row for clarity.}
    \label{tab:retrieved_snr20_h2o}
\end{table}

Figure~\ref{fig:corner} presents four corner plots investigating the 0.74 {\microns} \ce{H2O} feature (Figures~\ref{fig:corner}a, \ref{fig:corner}b) and the 0.9 {\microns} \ce{H2O} feature (Figures~\ref{fig:corner}c, \ref{fig:corner}d) for a quiet G2 star (Figures~\ref{fig:corner}a, \ref{fig:corner}c) and an extremely active G2 star (Figures~\ref{fig:corner}b, \ref{fig:corner}d). Table~\ref{tab:retrieved_snr20_h2o} presents the retrieved parameter values for every parameter at the same wavelength ranges as Figure~\ref{fig:corner}, including the upper and lower credible region envelope. The true values are provided in bold for clarity. In Figure~\ref{fig:corner}d, the 1D marginalized posterior distribution for $\mathrm{A_{s}}$ shows that the retrieved value is well below the true value, at 0.19, compared with Figure~\ref{fig:corner}c, wherein the retrieved $\mathrm{A_{s}}$ value is in line with the true value at 0.29. However, Figures~\ref{fig:corner}c and \ref{fig:corner}d are at the 0.9 {\microns} feature, which has known degeneracies with \ce{H2O} abundance and pressure. This can be seen in both corner plots, where the retrieved \ce{H2O} value is higher than the true value (-2.52) at -1.86 (Figure~\ref{fig:corner}c) and -1.75 (Figure~\ref{fig:corner}d), and the retrieved pressure is lower than the truth (0), for both the quiet (-0.32) and extreme (-0.39) stellar activity levels. This degeneracy is further described in BARBIE1, and can be broken by observing at shorter wavelengths. 

Figures~\ref{fig:corner}a and \ref{fig:corner}b show corner plots for the retrievals centered on the 0.74 {\microns} feature. In this bandpass, the pressure and \ce{H2O} retrieved value distributions are centered on the true values shown in Table~\ref{tab:retrieved_snr20_h2o}, with a relatively tight constraint. However, where in Figure~\ref{fig:corner}a the retrieved $\mathrm{A_{s}}$ value is slightly under the true value (0.24 compared to 0.3, respectively, although the true value is still firmly within the 68\% credible region), the retrieved $\mathrm{A_{s}}$ values in Figure~\ref{fig:corner}b are well below the true value, at 0.19. Looking at all of the 1D marginalized posterior distributions in Figures~\ref{fig:corner}a and \ref{fig:corner}b, we see that the posterior distributions are nearly identical for all parameters except $\mathrm{A_{s}}$.

There exists a clear degeneracy between extreme stellar activity and $\mathrm{A_{s}}$. Similarly, Figures~\ref{fig:corner}c and \ref{fig:corner}d show that the cloud posteriors differ significantly. We have assumed zero cloud coverage in this simulation, and as this is the edge of the grid space, we would expect a slightly over-retrieved value of clouds (as seen in Figure~\ref{fig:corner}c). However, Figure~\ref{fig:corner}d shows the cloud value is significantly over-retrieved at 15\% cloud coverage. We note that our analysis does not retrieve on planetary radius or other such bulk parameters; the perceived degeneracy with $\mathrm{A_{s}}$ due to the extreme stellar activity would manifest itself as uncertainty in the planet's radius and potentially phase angle. Stellar activity adds another potential source of degeneracy to all of these parameters.

\section{Discussion}
\label{sec:discuss}

\subsection{Implications for Spectroscopic Retrievals}

While the effects of stellar activity on retrievals of planetary atmospheres in transit and emission spectra are well-known and can be significant \cite[e.g.][]{wilson12,mccullough14,zellem17}, the effects on reflection spectra have been less studied. Efforts to study these effects are complicated by the fact that exoplanet host stars may vary widely in spectral type, and their activity levels can also vary widely within the same spectral type. Even so, most stars have activity levels low enough that their effect on retrievals is unlikely to pose a problem for a survey such as the Habitable Worlds Observatory. Typical field stars, such as those that are likely to be targets in an HWO-style survey, are only slightly more active than our Sun on average \cite{basri13}, and our results in Figures \ref{fig:heatmaps_h2o} and \ref{fig:corneractive} show that the differences between retrievals for such an active star and a quiet star to be negligible. Given this, our results show that if the effects of stellar activity on the data quality and speckle pattern are also negligible, then typical stellar activity will not have a significant impact on atmospheric retrievals from planetary albedo spectra.

In contrast to these Solar analogs, extremely active, rapidly-rotating stars can have much greater starspot coverage (see Figure \ref{fig:starspots}). This greater coverage has two significant effects on retrievals. First, the overall reduction in flux results in an under-retrieval of the planet's continuum albedo because the observed planet-to-star contrast ratio is smaller. This would result in an under-retrieval of the planet's radius and would contribute to other parameters that are degenerate with radius such as phase angle and cloud albedo. This effect would apply equally regardless of the bandpass or other parameters of the simulated observations.

Second, we suggested in Section \ref{sec:intro} a possible second-order effect. A large starspot coverage could result in errors or uncertainties in the stellar spectrum model due to contamination from \ce{H2O} molecules in the starspot spectrum itself, which could change the retrieved abundance of \ce{H2O} in the planet's atmosphere. We would expect that this effect could be mitigated by the choice of observational parameters (such as bandpass and SNR), given the much higher temperature of the stellar atmosphere, which will result in a different spectral shape for the same abundance---our results in Figure \ref{fig:heatmaps_h2o} bear this out. The planetary \ce{H2O} abundance can be retrieved for both quiescent and extremely active host stars in the 0.74 ${\rm \mu m}$ feature, and the increase in signal-to-noise ratio needed to do so is not onerous (SNR=18, as opposed to 13 for less active stars). Notably, this result is also independent of stellar spectral type. Although G stars and M stars have significantly different \ce{H2O} abundances in their quiescent spectra, we find no significant difference in any of our retrieval results between the two spectral types.

\subsection{Implications for Survey Design}
\label{sec:design}

The effects of stellar activity on retrievals are of particular concern when designing an HWO-style survey for potentially-habitable planets. The science requirements for such a survey apply additional observational constraints, which will need to be balanced against any constraints stemming from stellar activity. However most active stars are solar analogues for which the impact of starspots on a survey will be negligible, and any problems can be effectively eliminated by target selection, excluding the small number of highly active stars from the survey. However, it is worth considering the effect if an extremely active star were to be included in the survey for some reason (e.g. to investigate a high-interest candidate like Proxima b).

An especially notable observational constraint for an HWO-style survey is the known degeneracy between \ce{H2O} abundance and pressure in the 0.9 ${\rm \mu m}$ feature (see BARBIE1). \ce{H2O} is far easier to detect at 0.9 ${\rm \mu m}$, as is shown in Figure \ref{fig:heatmaps_h2o}, requiring a signal-to-noise ratio of only 5 compared with SNR=13 at 0.74 ${\rm \mu m}$. However, retrieving the abundance of \ce{H2O} accurately is much more limited at this wavelength. This would pressure a survey for exoEarths to focus on the 0.74 ${\rm \mu m}$ feature where this degeneracy can be broken. In this case, the added constraints from stellar activity are complementary, as 0.74 ${\rm \mu m}$ is also the wavelength needed to retrieval the \ce{H2O} abundance accurately in the presence of starspots. However, with the greater SNR requirement for extremely active stars, such a target would require perhaps twice as much observing time as a less active star. Thus, it is feasible to include extremely active targets in an HWO-style survey if necessary, but their number should likely be limited.

Given the increased challenges of retrieving on planets of highly active (or more precisely, highly variable) stars, a survey for exoEarths may want to deprioritize or even exclude these stars as targets. This may be a concern if the cutoff excludes a large fraction of the target list, but such extreme activity to too rare to be a serious concern. The impact of such a cutoff on a survey may be estimated by comparing a catalog of stellar variability with an example HWO target list such as that used by \cite{howe24}. That work used a target list of 168 stars generated by the AYO tool \cite{Stark2019} for a fiducial HWO observatory design utilizing an EMCCD detector. For observations of stellar variability, \cite{Fetherolf2023} produced a catalog of variability parameters for over 80000 \textit{TESS} targets, which include a representative subsample of 71 out of 168 stars on the \cite{howe24} target list. Of those 71 stars, 3 of them have observed variability $>1\%$, with the largest being $2.5\%$. (This included 7 out of 20 M-dwarfs on the target list ranging from M0 to M3, none of which showed variability $>1\%$.) Thus, even a very conservative target list that excludes all stars with variability $>1\%$ would only remove $\sim$4\% of likely HWO targets.

Similar results have been found for the \textit{Kepler} catalog. \cite{Basri2018} found that out of a large sample of \textit{Kepler} lightcurves of FGK stars, the vast majority had variability $<3\%$, and most slow rotators with rotation periods comparable to the Sun had variability $<1\%$. \cite{Rebull2017}, examining \textit{K2} lightcurves in the younger ($\sim$790 Myr) Praesepe cluster, found similar levels of variability among FGK stars there, but higher variability among M-dwarfs, up to $\sim10\%$. This activity level may pose challenges for surveys targeting younger targets order later-type M-dwarfs, but it is unlikely to seriously affect the search for exoEarths.



A simple cutoff in variability leaves the problem of a star that appears to be quiescent, but is in fact highly active with large spots not visible to the observer, which may contamination in the observed planetary spectrum, as we model in the ``worst-case scenarios'' in this work. However, this remains an unlikely scenario because of the system geometry required for it to happen. It can occur only for a highly-active star viewed nearly-pole on, with the spot or spots entirely in the far hemisphere. With only 4\% of likely target stars exhibiting high variability at all, the rate of targets with high undetected variability will be $<1\%$, and perhaps much lower. In all other cases, any spots visible to the observatory could be detected either by time variability, as in the \textit{Kepler} and \textit{TESS} lightcurves or, if large enough to be major contaminants, by peculiar spectral features.

We can also expect that any large starspots would almost certainly be visible to the observatory for host stars of planets discovered by the transit method. Thus, if any transiting planets are found that would be amenable to direct imaging characterization, they would be even safer targets for inclusion in a direct imaging survey, with only a particularly unusual spot geometry posing problems for them. 


\subsection{Non-Spectroscopic Effects and Future Work}

We note that stellar activity can have other effects on direct imaging in addition to those on the planetary spectrum, especially with regard to coronography. Although the stellar disk is not resolved, starspots and faculae create a spatially and spectrally varying stellar surface, which can affect the speckle pattern in the coronagraph. These features can also evolve with time (and stellar rotation), which can further change the speckle pattern over time, complicating the speckle subtraction. A fuller treatment of the effects of stellar activity on a future survey such as HWO will require an analysis of these effects on coronography.

We also do not consider stochastic variation in the observations. Retrieval results will be sensitive to the specific noise draw, and a different noise draw may result in slightly different SNR thresholds. These effects will be small, perhaps shifting the SNR thresholds by $\pm1$. However, we leave an in-depth analysis of stochastic noise effects to future work.

This analysis may also require consideration of shorter-period activity including faculae, flares, the stochastic variability of granulation, and differences in granulation between spectral types. Faculae in particular have a similar effect to starspots in the opposite direction, being hotter than the surrounding star. However, their effect is likely to be smaller than that for starspots. For our Sun, for example, the contribution of faculae to variability in total solar irradiance is smaller than sunspots by a factor of $\sim5$.

Likewise, longer-period time variability due to stellar rotation or changes in stellar activity may need to be taken into account. In particular, stellar time-variability could introduce an additional degeneracy with surface variation on the planet. A more complete analysis at the system level might also include close-in planets as well as exozodiacal dust.

We leave an analysis of these effects to future work, partly in anticipation of incorporating these new activity features into ExoVista, and future studies of coronography and other starlight suppression technologies should also incorporate these features.







\section{Conclusion}
\label{sec:conclusion}

Stellar activity is a concern for the detection and characterization of extrasolar planets. Some impacts of stellar activity such as those on transiting planets have been well studied. However, because future searches for potentially habitable planets will likely involve direct imaging of terrestrial-planets in reflection, a clear understanding of the uncertainties such as those introduced by stellar activity on these observations is needed.

We have use simulated spectroscopic retrievals to investigate one key source of this uncertainty across the parameter space of potentially habitable planet targets, wherein the observed spectrum of the star is different from the spectrum reflected by the planet due to unseen starspots. We find that unseen starspots will reduce the flux reflected by the planet, contributing to the existing uncertainty in the planet's radius. Water vapor features in the starspot's spectrum could potentially be a contaminant in measuring the abundance of water in the planet's atmosphere, but our results show that this is not a concern, requiring only a modestly higher signal-to-noise ratio to disentangle the signals, even for the most extreme stellar activity scenarios.

We also consider the impact of these effects on future surveys such as the proposed Habitable Worlds Observatory. In practice, the problematic cases are few in number and can usually be removed in the target selection phase by introducing a cutoff in photometric variability. Indeed, our results show no statistically significant error contribution from stellar variability as high as 1\%, which clears $\sim95\%$ of the parameter space of an HWO target list such as the one considered in \cite{howe24}, both for FGK stars and M-dwarfs. Only a small number of extreme cases are likely to cause problems, and only a fraction of those will have unfavorable system geometries that would result in such activity going undetected.
Thus starspot contaminants in planetary reflection spectra will not be a significant source of error in the search for potentially habitable planets. Future work should focus on a more complete survey of the variability of potential target stars, as well as studies of the effects of starspots on coronography to ensure that speckle subtraction can be carried out to the required precision for active stars.

\section*{Disclosures}

The authors have no conflicts of interest to declare.

\section*{Code and Data Availability}

PSGNest is available at https://psg.gsfc.nasa.gov/apps/psgnest.php.

ExoVista is available at https://github.com/alexrhowe/ExoVista.

\section*{Acknowledgments}

NL gratefully acknowledges financial support from a NASA FINESST and an appointment to the NASA Postdoctoral Program (NPP) at the NASA Goddard Space Flight Center, administered by Oak Ridge Associated Universities under contract with NASA.

ARH acknowledges support by NASA under award number 80GSFC24M0006 through the CRESST II cooperative agreement, as well as support from the GSFC Exoplanet Spectroscopy Technologies (ExoSpec).

This work was performed in part by members of the Virtual Planetary Laboratory Team, a member of the NASA Nexus for Exoplanet System Science, funded via NASA Astrobiology Program Grant No. 80NSSC18K0829.

We thank the anonymous referees whose comments have helped us to greatly improve the quality of this paper.

\bibliography{refs}

\begin{thebibliography}{10}

\bibitem{desort07}
M.~{Desort}, A.~M. {Lagrange}, F.~{Galland}, {\em et~al.}, ``{Search for exoplanets with the radial-velocity technique: quantitative diagnostics of stellar activity},'' {\em \aap} {\bf 473}, 983--993  (2007).

\bibitem{simola22}
U.~{Simola}, A.~{Bonfanti}, X.~{Dumusque}, {\em et~al.}, ``{Accounting for stellar activity signals in radial-velocity data by using change point detection techniques},'' {\em \aap} {\bf 664}, A127  (2022).

\bibitem{basri13}
G.~{Basri}, L.~M. {Walkowicz}, and A.~{Reiners}, ``{Comparison of Kepler Photometric Variability with the Sun on Different Timescales},'' {\em \apj} {\bf 769}, 37  (2013).

\bibitem{wilson12}
R.~E. {Wilson}, ``{Spotted Star Light Curves with Enhanced Precision},'' {\em \aj} {\bf 144}, 73  (2012).

\bibitem{mazeh15}
T.~{Mazeh}, T.~{Holczer}, and A.~{Shporer}, ``{Time Variation of Kepler Transits Induced By Stellar Rotating Spots{\textemdash}a Way to Distinguish between Prograde and Retrograde Motion. I. Theory},'' {\em \apj} {\bf 800}, 142  (2015).

\bibitem{Espinoza2019}
N.~{Espinoza}, B.~V. {Rackham}, A.~{Jord{\'a}n}, {\em et~al.}, ``{ACCESS: a featureless optical transmission spectrum for WASP-19b from Magellan/IMACS},'' {\em \mnras} {\bf 482}, 2065--2087  (2019).

\bibitem{Moran2023}
S.~E. {Moran}, K.~B. {Stevenson}, D.~K. {Sing}, {\em et~al.}, ``{High Tide or Riptide on the Cosmic Shoreline? A Water-rich Atmosphere or Stellar Contamination for the Warm Super-Earth GJ 486b from JWST Observations},'' {\em \apjl} {\bf 948}, L11  (2023).

\bibitem{FT2025}
M.~{Fournier-Tondreau}, Y.~{Pan}, K.~{Morel}, {\em et~al.}, ``{Transmission spectroscopy of WASP-52 b with JWST NIRISS: water and helium atmospheric absorption, alongside prominent star-spot crossings},'' {\em \mnras} {\bf 539}, 422--438  (2025).

\bibitem{mccullough14}
P.~R. {McCullough}, N.~{Crouzet}, D.~{Deming}, {\em et~al.}, ``{Water Vapor in the Spectrum of the Extrasolar Planet HD 189733b. I. The Transit},'' {\em \apj} {\bf 791}, 55  (2014).

\bibitem{zellem17}
R.~T. {Zellem}, M.~R. {Swain}, G.~{Roudier}, {\em et~al.}, ``{Forecasting the Impact of Stellar Activity on Transiting Exoplanet Spectra},'' {\em \apj} {\bf 844}, 27  (2017).

\bibitem{rackham24}
B.~V. {Rackham} and J.~{de Wit}, ``{Toward Robust Corrections for Stellar Contamination in JWST Exoplanet Transmission Spectra},'' {\em \aj} {\bf 168}, 82  (2024).

\bibitem{Espinoza2025}
N.~{Espinoza}, N.~H. {Allen}, A.~{Glidden}, {\em et~al.}, ``{JWST-TST DREAMS: NIRSpec/PRISM Transmission Spectroscopy of the Habitable Zone Planet TRAPPIST-1 e},'' {\em \apjl} {\bf 990}, L52  (2025).

\bibitem{roettenbacher11}
R.~M. {Roettenbacher}, R.~O. {Harmon}, N.~{Vutisalchavakul}, {\em et~al.}, ``{A Study of Differential Rotation on II Pegasi via Photometric Starspot Imaging},'' {\em \aj} {\bf 141}, 138  (2011).

\bibitem{Deming2025}
D.~{Deming}, M.~H. {Currie}, V.~S. {Meadows}, {\em et~al.}, ``{Minimizing Star-spot Contamination of Exoplanet Transit Spectroscopy Using Alternate Normalization},'' {\em \aj} {\bf 170}, 11  (2025).

\bibitem{stark22}
C.~C. {Stark}, ``{ExoVista: A Suite of Planetary System Models for Exoplanet Studies},'' {\em \aj} {\bf 163}, 105  (2022).

\bibitem{howe24}
A.~R. {Howe}, C.~C. {Stark}, and J.~E. {Sadleir}, ``{Scientific impact of a noiseless energy-resolving detector for a future exoplanet-imaging mission},'' {\em Journal of Astronomical Telescopes, Instruments, and Systems} {\bf 10}, 025008  (2024).

\bibitem{barbie1}
N.~{Latouf}, A.~M. {Mandell}, G.~L. {Villanueva}, {\em et~al.}, ``{Bayesian Analysis for Remote Biosignature Identification on exoEarths (BARBIE). I. Using Grid-based Nested Sampling in Coronagraphy Observation Simulations for H$_{2}$O},'' {\em \aj} {\bf 166}, 129  (2023).

\bibitem{barbie2}
N.~{Latouf}, A.~M. {Mandell}, G.~L. {Villanueva}, {\em et~al.}, ``{Bayesian Analysis for Remote Biosignature Identification on exoEarths (BARBIE). II. Using Grid-based Nested Sampling in Coronagraphy Observation Simulations for O$_{2}$ and O$_{3}$},'' {\em \aj} {\bf 167}, 27  (2024).

\bibitem{barbie3}
N.~{Latouf}, M.~D. {Himes}, A.~M. {Mandell}, {\em et~al.}, ``{BARBIE. Bayesian Analysis for Remote Biosignature Identification on exoEarths. III. Introducing the KEN},'' {\em \aj} {\bf 169}, 50  (2025).

\bibitem{vspec}
T.~M. {Johnson}, C.~{Kelahan}, A.~M. {Mandell}, {\em et~al.}, ``{VSPEC: Variable Star PhasE Curve}.'' Astrophysics Source Code Library, record ascl:2504.013  (2025).

\bibitem{magic2014}
Z.~{Magic} and M.~{Asplund}, ``{The Stagger-grid: A grid of 3D stellar atmosphere models - VI. Surface appearance of stellar granulation},'' {\em arXiv e-prints} , arXiv:1405.7628  (2014).

\bibitem{Kurucz03}
F.~{Castelli} and R.~L. {Kurucz}, ``{New Grids of ATLAS9 Model Atmospheres},'' in {\em Modelling of Stellar Atmospheres},  N.~{Piskunov}, W.~W. {Weiss}, and D.~F. {Gray}, Eds.,  {\bf 210}, A20  (2003).

\bibitem{Husser13}
T.~O. {Husser}, S.~{Wende-von Berg}, S.~{Dreizler}, {\em et~al.}, ``{A new extensive library of PHOENIX stellar atmospheres and synthetic spectra},'' {\em \aap} {\bf 553}, A6  (2013).

\bibitem{Kopal1950}
Z.~{Kopal}, ``{Detailed effects of limb darkening upon light and velocity curves of close binary systems},'' {\em Harvard College Observatory Circular} {\bf 454}, 1--12  (1950).

\bibitem{Claret2013}
A.~{Claret}, P.~H. {Hauschildt}, and S.~{Witte}, ``{New limb-darkening coefficients for Phoenix/1d model atmospheres. II. Calculations for 5000 K {\ensuremath{\leq}} T$_{eff}$ {\ensuremath{\leq}} 10 000 K Kepler, CoRot, Spitzer, uvby, UBVRIJHK, Sloan, and 2MASS photometric systems},'' {\em \aap} {\bf 552}, A16  (2013).

\bibitem{Claret2012}
A.~{Claret}, P.~H. {Hauschildt}, and S.~{Witte}, ``{New limb-darkening coefficients for PHOENIX/1D model atmospheres. I. Calculations for 1500 K {\ensuremath{\leq}} T$_{eff}$ {\ensuremath{\leq}} 4800 K Kepler, CoRot, Spitzer, uvby, UBVRIJHK, Sloan, and 2MASS photometric systems},'' {\em \aap} {\bf 546}, A14  (2012).

\bibitem{ChenKipping}
J.~{Chen} and D.~{Kipping}, ``{Probabilistic Forecasting of the Masses and Radii of Other Worlds},'' {\em \apj} {\bf 834}, 17  (2017).

\bibitem{PSG}
G.~L. {Villanueva}, M.~D. {Smith}, S.~{Protopapa}, {\em et~al.}, ``{Planetary Spectrum Generator: An accurate online radiative transfer suite for atmospheres, comets, small bodies and exoplanets},'' {\em \jqsrt} {\bf 217}, 86--104  (2018).

\bibitem{PSGbook}
G.~L. {Villanueva}, G.~{Liuzzi}, S.~{Faggi}, {\em et~al.}, {\em {Fundamentals of the Planetary Spectrum Generator}}, Self-Published  (2022).

\bibitem{feng18}
Y.~K. {Feng}, T.~D. {Robinson}, J.~J. {Fortney}, {\em et~al.}, ``{Characterizing Earth Analogs in Reflected Light: Atmospheric Retrieval Studies for Future Space Telescopes},'' {\em \aj} {\bf 155}, 200  (2018).

\bibitem{luvoir}
{The LUVOIR Team}, ``{The LUVOIR Mission Concept Study Final Report},'' {\em arXiv e-prints} , arXiv:1912.06219  (2019).

\bibitem{multinest}
F.~{Feroz}, M.~P. {Hobson}, and M.~{Bridges}, ``{MULTINEST: an efficient and robust Bayesian inference tool for cosmology and particle physics},'' {\em \mnras} {\bf 398}, 1601--1614  (2009).

\bibitem{benneke13}
B.~{Benneke} and S.~{Seager}, ``{How to Distinguish between Cloudy Mini-Neptunes and Water/Volatile-dominated Super-Earths},'' {\em \apj} {\bf 778}, 153  (2013).

\bibitem{harrington22}
J.~{Harrington}, M.~D. {Himes}, P.~E. {Cubillos}, {\em et~al.}, ``{An Open-source Bayesian Atmospheric Radiative Transfer (BART) Code. I. Design, Tests, and Application to Exoplanet HD 189733b},'' {\em The Planetary Science Journal} {\bf 3}, 80  (2022).

\bibitem{MamajekII}
M.~J. {Pecaut} and E.~E. {Mamajek}, ``{Intrinsic Colors, Temperatures, and Bolometric Corrections of Pre-main-sequence Stars},'' {\em \apjs} {\bf 208}, 9  (2013).

\bibitem{hathaway15}
D.~H. {Hathaway}, ``{The Solar Cycle},'' {\em Living Reviews in Solar Physics} {\bf 12}, 4  (2015).

\bibitem{harre21}
J.-V. {Harre} and R.~{Heller}, ``{Digital color codes of stars},'' {\em Astronomische Nachrichten} {\bf 342}, 578--587  (2021).

\bibitem{Stark2019}
C.~C. {Stark}, R.~{Belikov}, M.~R. {Bolcar}, {\em et~al.}, ``{ExoEarth yield landscape for future direct imaging space telescopes},'' {\em Journal of Astronomical Telescopes, Instruments, and Systems} {\bf 5}, 024009  (2019).

\bibitem{Fetherolf2023}
T.~{Fetherolf}, J.~{Pepper}, E.~{Simpson}, {\em et~al.}, ``{Variability Catalog of Stars Observed during the TESS Prime Mission},'' {\em \apjs} {\bf 268}, 4  (2023).

\bibitem{Basri2018}
G.~{Basri} and H.~T. {Nguyen}, ``{Double Dipping: A New Relation between Stellar Rotation and Starspot Activity},'' {\em \apj} {\bf 863}, 190  (2018).

\bibitem{Rebull2017}
L.~M. {Rebull}, J.~R. {Stauffer}, L.~A. {Hillenbrand}, {\em et~al.}, ``{Rotation of Late-type Stars in Praesepe with K2},'' {\em \apj} {\bf 839}, 92  (2017).

\end{thebibliography}
\bibliographystyle{spiejour}   

\end{document}